**Recyclable thin-film soft electronics for smart packaging and e-skins**

*Manuel Reis Carneiro, Aníbal T. de Almeida, Mahmoud Tavakoli\* and Carmel Majidi\**

M. Reis Carneiro, Professor C. Majidi\*

Soft Machines Lab, Department of Mechanical Engineering, Carnegie Mellon University, Pittsburgh, Pennsylvania 15213, United States

E-mail: cmajidi@andrew.cmu.edu

M. Reis Carneiro, Professor Aníbal T. de Almeida, Professor M. Tavakoli\*

Institute of Systems and Robotics, Department of Electrical and Computer Engineering, University of Coimbra, Coimbra 3030-290, Portugal

E-mail: mahmoud@isr.uc.pt

Abstract: Despite advances in soft, sticker-like electronics, few efforts have dealt with the challenge of electronic waste. Here, we address this by introducing an eco-friendly conductive ink for thin-film circuitry composed of silver flakes (Ag) and a water-based polyurethane (WPU) dispersion. This ink uniquely combines high electrical conductivity ($1.6 \times 10^5$ S/m), high resolution digital printability, robust adhesion for microchip integration, mechanical resilience, and recyclability. Recycling is achieved with an ecologically-friendly processing method to decompose the circuits into constituent elements and recover the conductive ink with a decrease of only 2.4% in conductivity. Moreover, adding liquid metal enables stretchability of up to 200% strain, although this introduces the need for more complex recycling steps. Lastly, we demonstrate on-skin electrophysiological monitoring biostickers along with a recyclable smart package with integrated sensors for monitoring safe storage of perishable foods.

1. Introduction

The accumulation of electronic waste (e-waste) is a pressing global problem that poses environmental threats to natural ecosystems, economic burdens related to the loss or insufficient recovery of valuable resources (e.g. precious metals), and health concerns due to the use and improper discarding of toxic substances[1,2]. This challenge is only expected to deepen as electronics become more ubiquitous through the use of wearable, printable, and sticker-like systems that are currently being developed for use in varied fields from healthcare to industry [3,4,13–18,5–12]. Like conventional electronics, flexible thin-film circuits could add to the growing e-waste problem since there are no established recycling methods for their unique materials and manufacturing processes.[19,20]. Moreover, unlike the traditional rigid electronics that are intended for long-term use, most thin-film soft electronics are being designed as disposable devices for use in health monitoring, IoT, and smart packaging[21–25], having the potential to dramatically increase the quantity of e-waste in the coming decades.

Another emerging environmental challenge is the use of toxic materials in the synthesis and decomposition of soft and wearable electronic systems. Current efforts typically rely on conductive inks and composites that are printed over polymeric substrates[26–28] and often rely on toxic organic solvents in to achieve adequate rheology for printing and processing[29]. If improperly handled or discarded, such solvents can impact human health and threaten natural ecosystems[19]. Furthermore, some of these e-inks have a limited shelf-life and must be stored and handled within a specific temperature range (usually with refrigeration), thereby further adding to their cost and environmental burden on account of increased energy consumption and pollution [30–35].

Here, we address these challenges by introducing a new class of printable conductive inks, and fabrication techniques that enable digital fabrication of sustainable, and eco-friendly soft electronics. This ink uniquely combines high resolution digital printing, microchip integration, strain tolerance, and simple, efficient and ecological recycling. Moreover, the microchip integration is performed without the addition of a separate electrically conductive adhesive (ECA), which is common in printed electronics[36–40], thus reducing the fabrication cost and complexity. Finally, all processing steps are performed in ambient conditions, with no need for thermal sintering of the printed circuit. This reduces energy consumption during fabrication and permits printing on a wide range of heat-sensitive substrates, including sustainable plastics that are of increasing interest for green electronic technologies such as paper[41,42], textiles[43], or polymers[44,45] and bio-synthetic composite materials[46–49].

These inks are composed of various combinations of waterborne polyurethane dispersion (WPU), Ag flakes, and liquid metal (LM) alloy and are free of organic solvents. Combinations of these materials result in printable inks that exhibit high electrical conductivity, tolerance to mechanical strain, robust adhesive properties, and sinter-free conductivity, making it compatible for printing on soft elastomeric thin films and other fragile heat-sensitive substrates. The same ink can also be used to attach surface mounted devices (SMDs) and other microelectronic components for creating complex thin-film circuits, thereby eliminating the need for additional soldering steps and materials. **Table S1** compares this formulation with other recent efforts on recyclable electronics. Despite promising advancements with these previous attempts, no one conductive ink has simultaneously satisfied all of the necessary requirements for scalable fabrication and large scale deployment. These include the ability to integrate surface mounted microelectronic chips or the ability to support digital printing, extreme mechanical deformation, and ecologically-friendly pathways for ink synthesis and decomposition.

One version of this ink is composed of a percolating network of microscale silver flakes (**Fig. 1A**). This water-based conductive polymer composite is recyclable, has a conductivity of ~1.1 - $1.6 \times 10^5$ S/m, and can be stored at room temperature for more than 4 weeks without impacting its conductivity or printability (**Figure 1B**). This eco-friendly ink can be printed digitally to create flexible and soft electronic circuits, which can be combined with rigid IC chips and SMD components through a "soft soldering" process at room temperature (Figure 1C). The printing scalability and SMD integration enable numerous applications, including ultra-flexible, surface-conformable, and recyclable circuits for consumer electronics as a substitute for flexible PCBs (fPCBs) and molded interconnect devices (MIDs), as well as a skin-conformable health monitoring sticker with integrated electronics for skin temperature monitoring and electrophysiology (Figure 1D).
Recycling of circuits is achieved through a facile process at room temperature, as depicted in Figure 1E. The process involves soaking circuits in isopropyl alcohol (IPA) to achieve decomposition and separation of the circuits into fundamental components (ink, polymeric substrate, and microchips), without requiring special equipment or toxic solvents. The recovered ink aggregates can then be washed with IPA to enable separation of the PU and silver flakes, which can be reused to synthesize new inks using the original method. This eco-friendly process facilitates proper disposal or recycling of the circuit components.

The proposed combination of materials and uncomplicated techniques results in sophisticated eco-friendly soft electronic circuits with a fully-circular life-cycle. These materials and methods encompass the creation of an environmentally friendly conductive polymer that can be easily utilized in digital printing processes, a simple and reliable approach to integrate SMD chips at room temperature without the need for sintering, and an inexpensive method for separating and breaking down the circuits, allowing for the retrieval and reuse of their components in the production of new circuits for diverse applications across various fields.

2. Results and Discussion

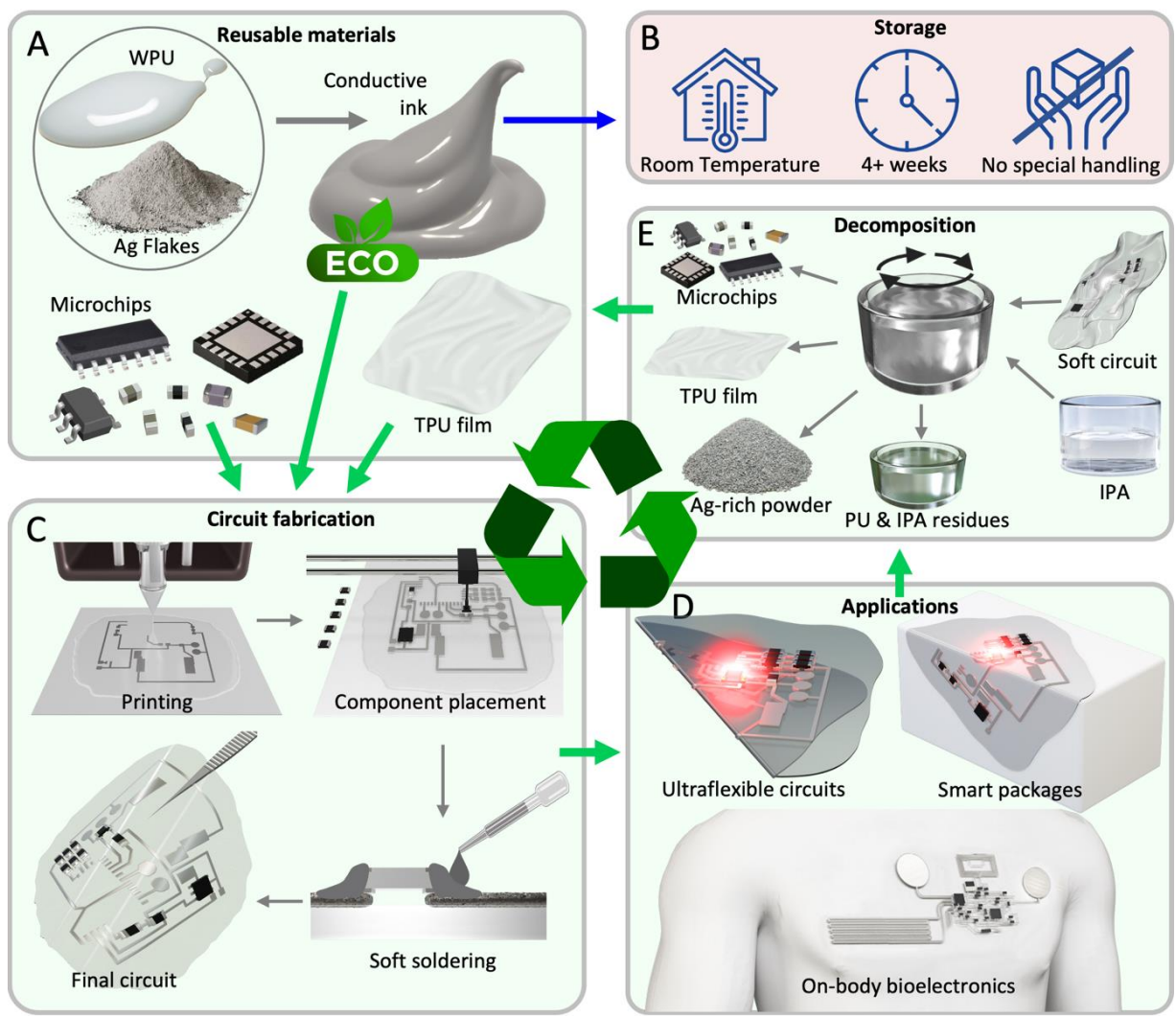

**Figure 1.** Green soft electronics with a circular life cycle. **A:** Eco-friendly water based conductive ink composed of silver flakes and polyurethane for wiring rigid microchips and surface mount devices on a polyurethane film substrate. **B:** The water-based Ag-WPU ink can be stored at room temperature for more than 4 weeks without impacting conductivity or printability. **C:** Room temperature circuit fabrication process includes digital circuit printing, microchip pick-and-placing, and electromechanical bonding of components through a soft 'soldering' process. **D:** Applications include ultraflexible circuits, compliant electronics, and skin-mounted bioelectronics. **E:** An organic solvent-free decomposition process based on the swelling of polyurethanes when in contact with Isopropyl Alcohol (IPA) enables separation of the soft circuits in their fundamental components that can be further recycled into a new ink or further reprocessed.

## 2.1. Ink synthesis and soft circuit fabrication

As depicted in **Figure 2A**, the ink is prepared by adding Ag flakes in an organic solvent-free waterborne polyurethane (WPU) dispersion. Deionized water is then added to tune the ink rheology so that it can be printed. The solution is further mixed, resulting in a silvery ink with paste-like consistency shown in **Figure 2B**.

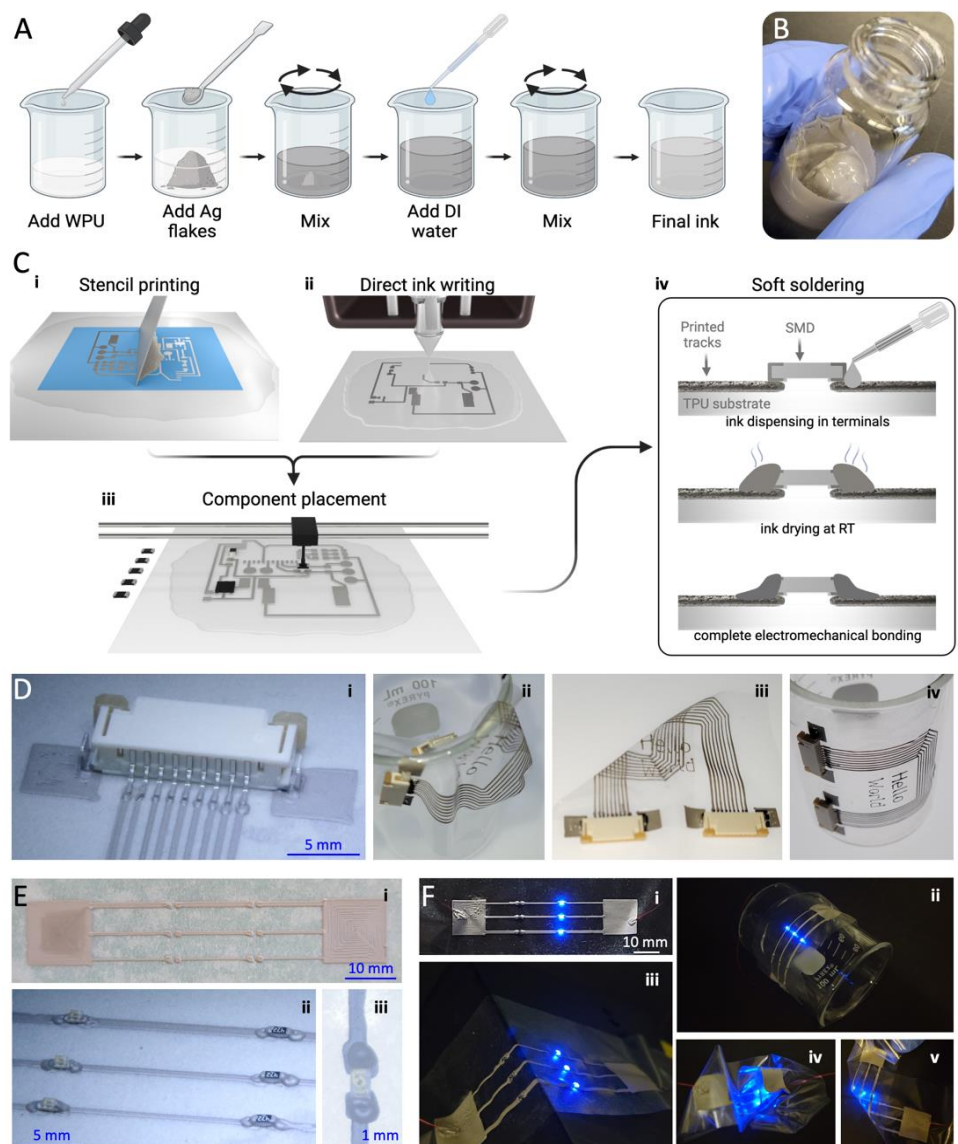

**Figure 2. A:** Synthesis of the sinter free water-based conductive ink. **B:** Synthetized Ag-WPU ink with paste-like consistency. **C:** Soft circuits fabrication process. Conductive lines are printed using (i) stencil printing or (ii) direct ink writing using a digital printer. (iii) Microchips and SMD components are placed on the printed circuit. (iv) Droplets of the Ag-WPU conductive compound are dispensed on the microchip terminals and dried at room temperature, forming a strong and reliable electromechanical bonding between the rigid component and the underlying conductive tracks. **D:** (i) Detail of an SMD flat cable connector bonded to a set of printed conductive lines (680 um pin separation). This circuit can work as a flexible multi-signal conductor which is (ii) ultraflexible, (iii) highly deformable, (iv) and can be molded to non-planar surfaces to work as a MID. **E:** (i) Example of digitally printed circuit traces (ii) populated with 0603 size SMD resistors and LEDs. Line width from top to bottom is 1000 um, 600 um and 400 um. (iii) Detail of the soft bonding points between a 0603 size LED and the underlying conductive tracks. **F:** (i) The circuit remains functional even when (ii, iii) conformed to non-planar surfaces or (iv, v) highly deformed, withstanding crumpling, as well as 90º and 180º bends.

Conductive traces and complete digital circuits can be patterned through stencil printing

(**Figure 2C i**) or direct ink writing (**Figure 2C ii**). While the use of a stencil allows for easy scalability of the printing procedure, direct ink writing allows for automated stencil-free additive fabrication of circuits with less ink waste. The ink is printed over a 50 μm thick polyurethane film and the printed ink is left to dry for 5 minutes to evaporate all water in the conductive paste. The full process is done at room temperature. To ensure the quality of printed or written ink on the substrate, special care was taken to prevent any interruptions or failures in the printed traces. Furthermore, post-printing, a visual inspection was conducted to guarantee a smooth surface, free from imperfections and discontinuities.

To implement hybrid circuits with integrated microchips, the rigid electronic components are placed over the printed circuit using either a pick-and-place system (**Figure 2C iii**) or manually with tweezers, as would be done for conventional rigid printed circuit boards (PCBs). The microchips are then integrated into the circuit through a room-temperature soft bonding process, as shown in **Figure 2C iv**. To achieve both mechanical bonding and electrical connection between the SMDs and the underlying printed conductive track, Ag-WPU ink droplets are dispensed on the SMD terminals using a syringe with a stainless steel tip (200 μm inner diameter) and dried for 5 minutes at room temperature. Through this "soft soldering" process, a conformal bond around the rigid component terminals and the underlying conductive track can be achieved.

In **Figure 2D i**, an SMD flat cable connector (680 μm pin separation) is shown bonded to a set of conductive lines printed through direct ink writing, and it can be observed that the polymeric soft solder joints conform well to the rigid terminals of the SMD component. This circuit can work as a flexible multi-signal conductor which is ultra-flexible, and capable of withstanding 180-degree bend with minimal (sub-mm) bending radius, as shown in **Figure 2D ii, iii**. Moreover, since the overall circuit is so thin (~100 μm) it can be molded to non-planar surfaces to work as a molded interconnect device (MID)[50]. For instance, in **Figure 2D iv**, the printed circuit is shown adhered to a curved glass surface only by means of the natural tackiness and stiction of the elastomer substrate.

**Figure 2E i** depicts a printed circuit with line widths of 1000 μm, 600 μm, and 400 μm respectively from top to bottom. In this circuit 0603-sized (1.5x0.8 mm package dimensions and 0.7 mm pad separation) SMD LEDs and resistors were bonded as shown in **Figure 2E ii**. A detail of the circuit showing the bonded LED and the printed lines and pads is shown in **Figure 2E iii**. The functioning circuit with illuminated LEDs is shown undeformed in **Figure 2F i**. Even when adhered to non-planar surfaces and when crumpled or bent (up to 180º bending with a sub-mm radius of curvature), the soft printed circuit with integrated SMD components remains functional as shown in **Figure 2F ii-v**. Integration of smaller SMD components in the printed soft circuits was also shown to be functional, as seen in **Figure S1** where 0402-sized resistors and LEDs (1.0x0.5 mm package dimensions and 0.5 mm pad separation) were bonded to a set of printed lines with 200 μm width which corresponds to the highest resolution print achieved through DIW. **Video S1** shows an ultraflexible printed circuit that is deformed (through bending and crumpling) without losing functionality.

**2.2. Characterization**

2.2.1. Conductivity and shelf life

As shown in **Figure 3A**, the conductivity of the Ag-WPU (89.2% Ag wt) printed trace is ~$1.16 \times 10^5$ S/m soon after the ink dries (day 0) and slightly improves over time reaching ~$1.54 \times 10^5$ S/m after 30 days of being printed and the printed lines being left unprotected at room temperature. This behavior can be explained by moisture slowly evaporating over time after the bulk $H_2O$ (the ink wet medium) is initially evaporated, which leads to an increase in the ratio of silver filler particles per volume of non-conductive PU, hence the higher conductivity. This behavior was previously observed in other conductive particle-filled polymeric compounds[51].

In addition, we conducted an investigation into the influence of silver concentration on the conductivity of ink. To achieve this, we varied the weight percentage (wt%) of Ag flakes relative to the solid contents of the WPU. Our findings align with existing research on percolative networks. We observed that when the silver concentration in the ink falls below 75%, the printed traces do not exhibit conductivity. However, as the concentration surpasses 90%, we noted a rapid increase in conductivity, as depicted in **Figure S2**. It is important to note that beyond 90% Ag wt%, the ink becomes excessively viscous, resembling a thick paste. This characteristic renders extrusion printing impractical and may also affect the mechanical properties of the ink due to the decreased polymer content. Nevertheless, we acknowledge the potential for further refinement of the ink formulation to achieve an optimal balance between higher electrical conductivity and rheology that suits DIW.

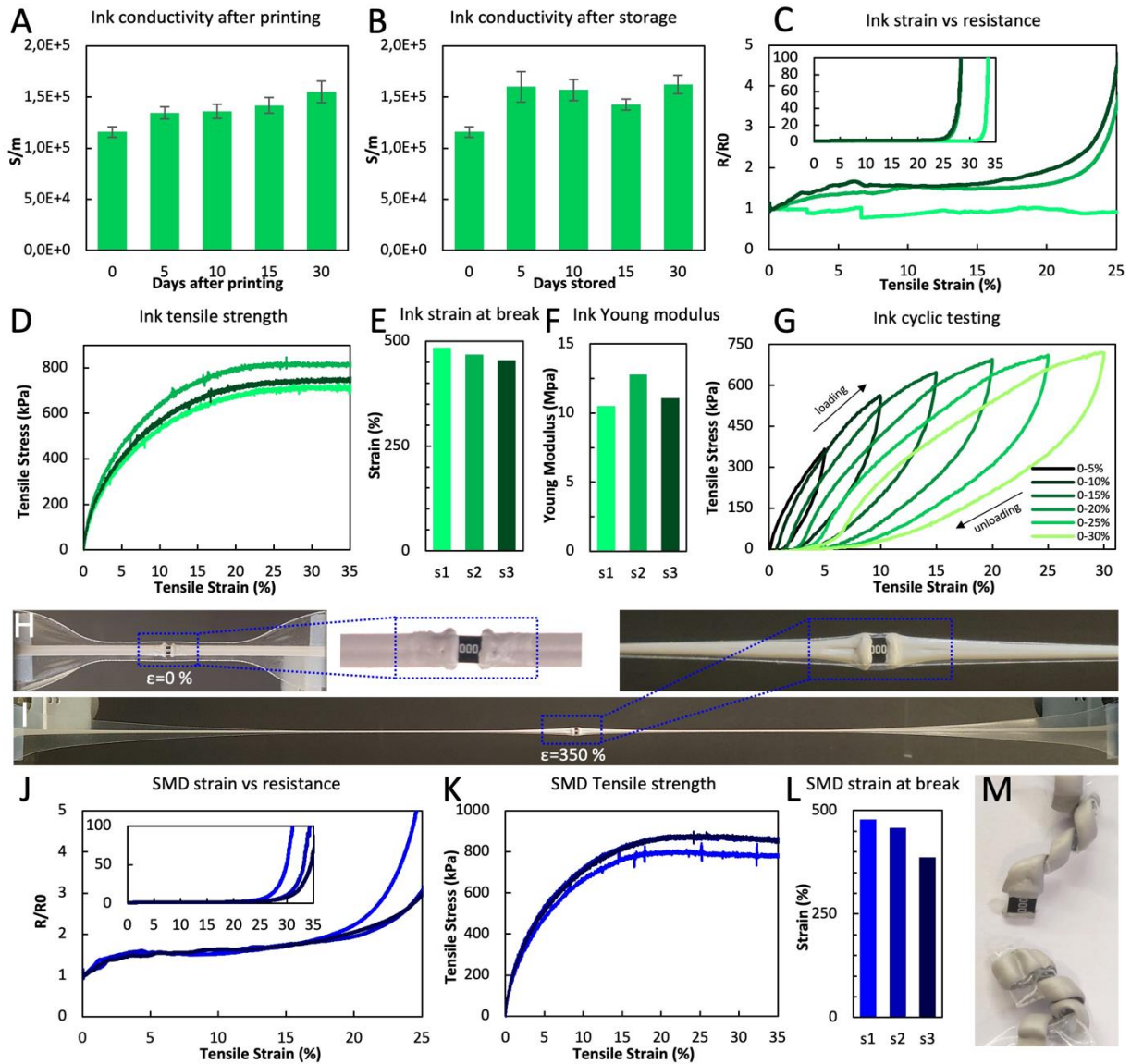

**Figure 3. A:** Ag-WPU conductivity and aging of exposed printed traces for up to 30 days. **B:** Ag-WPU conductivity for ink vials stored up to 30 days at room temperature. Error bars represent standard deviation **C:** Strain vs resistance curve for Ag-WPU traces printed over a thermoplastic polyurethane (TPU) substrate (three samples from distinct ink batches). **D:** Strain (0-30%) vs stress plot for the three printed ink samples. **E:** strain at break for the three ink samples. **F:** Estimated Young modulus for the 3 ink samples. **G:** Cyclic test of the Ag-WPU ink. **H:** Printed Ag-WPU track with integrated 0 Ω SMD resistor at 0% strain. **I:** Printed Ag-WPU track with integrated 0 Ω SMD resistor at 350% strain, before mechanical failure (electrical failure had already occurred). **J:** Strain vs resistance curve for Ag-WPU traces printed over TPU substrate with integrated 0 Ω SMD resistors (three samples from distinct ink batches). **K:** Strain (0-30%) vs stress plot for the three printed ink samples with integrated resistors. **L:** Strain at break for the three ink samples with integrated SMD resistors. **M:** Mechanical fracture of the samples occurs at the interface between soft printed lines and the rigid SMD component.

A batch of ink was fabricated, divided into closed vials, and stored at room temperature without special handling or storing precautions. The vials were exposed to sunlight through windows and artificial light on a benchtop to test their shelf life. As observed in **Figure 3B**, the conductivity of the aged ink lies between $1.16 \times 10^5$ S/m for the freshly synthesized ink, and $1.62 \times 10^5$ S/m for the ink stored for 30 days. Conductivity increase is due to water evaporation during storage, which increases the Ag flake concentration.

Viscosity also increases over time, but it does not affect printability for the first 30 days. Figure S3 shows the relationship between the viscosities of four Ag-WPU samples that were stored at room temperature in closed vials for different periods (0, 15, 30 and 35 days) at different shear rates. Although in the first 30 days, the viscosity of the inks at a $0.1$ s$^{-1}$ increases slightly from 125.65 to 220.91 Pa·s, the ink is still printable both by stencil printing and DIW. However, between 30 and 35 days of storage the viscosity increases rapidly to 789.79 Pa·s due to the curing of the polyurethane caused by low water content, rendering the ink unusable as it is too viscous for reliable printing. Lower temperature storage can address this issue by reducing water evaporation and keeping the ink's viscosity stable for longer periods. In addition, all four samples exhibit shear thinning behavior which proves beneficial in DIW."

2.2.2. Electromechanical characteristics of Ag-WPU ink

As observed in Figure 3C, the normalized resistance ($R/R_0$) for three Ag-WPU printed traces remains below 2 Ω/Ω when the sample is stretched up to ~22%, for two of the samples, while the third sample shows a normalized resistance below 2 Ω/Ω up to ~30% strain. At larger strains, the resistance increases rapidly for all three samples being that two of the samples fail at 28.3% strain and one of them at 33.5% strain. Electrical failure was considered to occur when the normalized resistance increased up to 100 Ω/Ω.

To further explain the visible difference between the 3 reported samples in Figure 3C, 6 more samples were fabricated from separate ink batches and tested in the same conditions. As shown in Figure S5, in 5 of the 9 Ag-WPU samples the normalized resistance ($R/R_0$) remains below 2 Ω/Ω at strains between 22 and 27%, while one of the samples shows $R/R_0 > 2$ Ω/Ω at strains higher than 15 %, and a last sample shows $R/R_0 > 2$ Ω/Ω only at strains larger than ~30%.

The observed variations among the samples are hypothesized to be a result of minor inconsistencies inherent to the sample preparation process. This encompasses a number of factors including inconsistencies in the preparation of the ink batches. The ink preparation process can be affected by variations in mixing or ambient room temperature and humidity, which can lead to disparities in the distribution of silver flakes within the polyurethane matrix.

In turn, these disparities could be a contributing factor to alterations in the percolation threshold. The percolation threshold - a pivotal factor in the composite's conductivity and strain sensitivity - is thus subjected to variation, culminating in the discernible differences in these properties among all samples. Moreover, the fabrication process of the dogbones introduces another layer of variability. The procedure entails printing on a thermoplastic polyurethane (TPU) substrate, a process that may introduce minor variations due to factors such as printing repeatability, the specific conditions of the printing process, or slight inconsistencies in the TPU substrate itself. Collectively, these factors might lead to additional variability in the final samples.

In terms of the mechanical properties of the ink, the strain stress curve between 0-35% strain for all three tested samples is shown **Figure 3D**. For all samples, the yield strength occurs at around 30% strain, with a magnitude of ~738-816 kPa. As shown in **Figure 3E**, the mechanical failure of the samples occurred between 453.8 and 484.5% strain (much above the electrical failure strain), and the young modulus of the Ag-WPU ink (plotted in **Figure 3F**) was estimated to be in the 10.5-12.8 MPa range.

To test the cyclic performance of the Ag-WPU composite, it was first stretched by 5% and then relaxed to its original length, and the strain was increased by 5% up to 30% for each successive cycle as shown in **Figure 3G**. While the maximum stress for each cycle follows the same values as for the strain-stress curve in **Figure 3D**, a hysteresis loop between loading and unloading cycles can be seen, which has been previously observed in other PU-based compounds[52]. As well, after 20% strain, some degree of plastic deformation can be observed, leading to a permanent stretch of ~3.5% in the sample for the following cycles.

The SEM images in **Figures S6 and S7**, corresponding to Ag-WPU show the structure and morphology of the ink, where only silver flakes are observable. Ag flakes have a disk-like shape with widths below 5 μm. In terms of orientation, the silver flakes appear randomly oriented in the sample, with no predominant alignment or preferred direction observed. The silver flakes are uniformly and evenly distributed across the entire sample, without visible clusters or void areas.

2.2.3. Electromechanical characterization of soft solder joints

A solid-state resistor was embedded in the printed traces using the proposed soft soldering method as depicted in **Figure 3H** and stretched (**Figure 3I**) until mechanical failure. As shown in **Figure 3J**, the normalized resistance for three printed samples with integrated SMD resistors remains below 2 Ω/Ω for strains of up to 20%, while full electrical failure (normalized resistance above 100 Ω/Ω) occurs above 31% strain for all three samples. The plot in **Figure**

**3K** provides evidence that the integration of the solid-state chip doesn't significantly affect the morphology of the strain-stress curve when compared with samples that do not contain an integrated resistor. Yield occurs at around 30 % strain, with a magnitude of ~775-854 MPa for the 3 tested samples. Furthermore, mechanical failure of all three samples occurred above 386.5 %, once again, much above their electrical failure, as plotted in **Figure 3L**. **Figure 3M** depicts the failure mode of the samples with integrated solid-state technology (SST) resistors. In all three samples, mechanical fracture occurred at the interface between the printed lines and the resistor. Nevertheless, no delamination of the Ag-WPU ink from the TPU substrate was ever observed, despite the occurrence of buckling in the printed ink layer due to plastic deformation at high strains. Plots of the full strain stress curves until fracture are provided in **Figure S8**. **Video S2** shows a printed circuit with integrated SMDs being stretched and crumpled without delamination of the rigid components or the conductive traces.

2.3. Circuit separation and recycling

One of the advantages of the Ag-WPU conductive polymer is that it can easily be degraded through a simple decomposition process and the surface mounted circuit components can be individually recovered. To do this, the soft circuits are soaked in an IPA-filled beaker, as shown in **Figure 4A i, ii**. The beaker is then placed on a magnetic stirrer and its content is stirred for 30 minutes at 500 RPM, as seen in **Figure 4A iii**. At this stage, one can observe that the solution turns gray due to the disintegration of the conductive ink from the substrate. As shown in **Figure 4A iv, v**, The TPU substrate as well as the rigid components can be removed from the beaker with just trace amounts of ink and the Ag flakes and conductive ink residues are left to precipitate for 1 hour, as shown in **Figure 4A vi**. The IPA can then be decanted and evaporated, leaving the ink residue and Ag flakes. Soaking of the circuits in water was also tested but no degradation was observed in this case.

The recycling process described above is based on the way IPA interacts differently with TPU and WPU due to their underlying chemical compositions and structure. Despite both being polyurethanes, TPU has a highly crosslinked structure compared to WPU and, when in contact with IPA, it can cause the TPU to swell as the alcohol molecules get lodged between the polymer chains, but it doesn't break down the actual chains. Once the alcohol evaporates, the TPU returns to its original state. WPU, on the other hand, is designed to be dispersible in water thanks to a less crosslinked structure and the presence of polar groups - hydrophilic segments - which are not as prevalent in TPU, making WPU more vulnerable to other solvents, specifically polar solvents such as IPA[53]. As such, when IPA comes into contact with WPU, it interacts with the polar soft segments breaking the polymer chains apart, causing the material to dissolve or degrade.

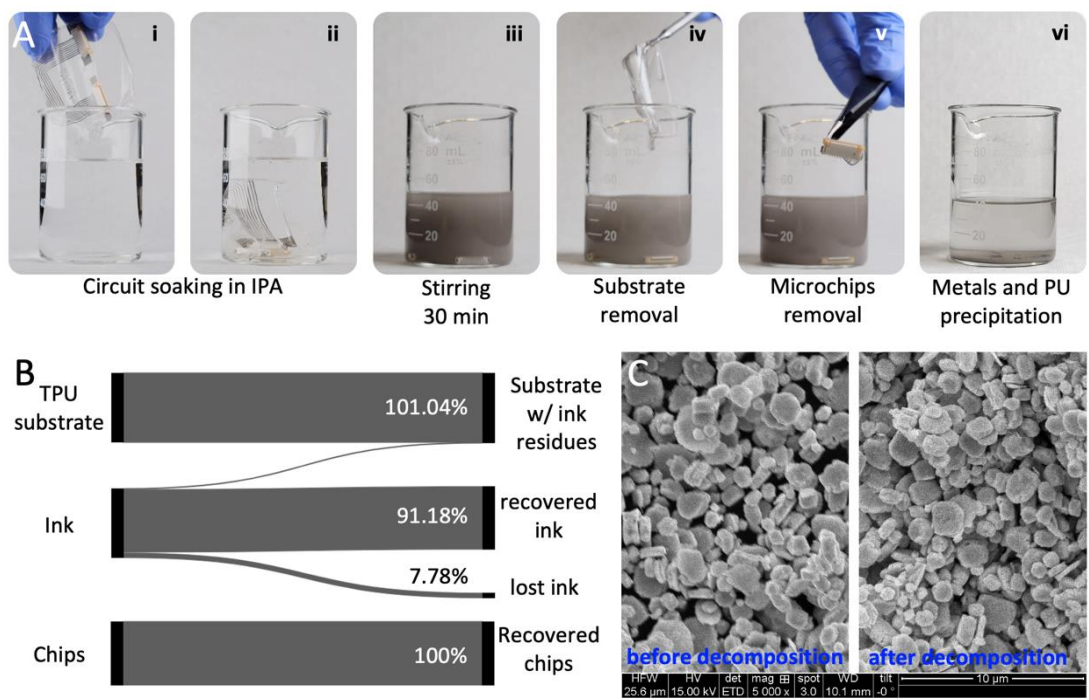

**Figure 4. A:** Circuit degradation and separation process. The circuit is soaked in IPA (i, ii) and stirred in a magnetic stirrer for 30minutes (iii). At this point, the clean TPU substrate (iv) and rigid components (v) can be removed from the solution. The suspended Ag flakes and PU residues are left to precipitate (vi) and are decanted, while any remaining IPA traces is then evaporated. **B:** Separation efficiency. While the rigid electronic components and TPU substrate can be fully recovered with trace amounts of ink residues, the ink (Ag flakes and PU residues) can be separated with ~90% efficiency due to losses during decanting. C: SEM image depicting no changes in morphology of the Ag flakes before and after the separation process (scale bar 10um)

As shown in **Figure 4B**, the process is efficient in enabling the recovery of all of the TPU substrate with trace amount of ink residue, as well as all the rigid SMD components that can be cleaned by spraying them with IPA above the recovery beaker. The precipitation, decantation, and IPA evaporation processes allow for ~91.18% of the initial ink (by weight) to be recovered, while ~7.78% of it is lost during processing (the remaining 1.04% is bonded to the TPU substrate). This separation efficiency estimate was obtained by comparing the weight of recovered components relative to the amount of components in the initial circuit. From the SEM image in **Figure 4C**, it can be observed that the morphology of the Ag flakes in the ink is not affected by the separation process, indicating that the flakes can be further recycled and reused. This circuit degradation and separation method is intended for separating the soft circuits into their basic components (shown in **Figure S9**) so that some components, such as the PU substrate, and rigid components can be directly reused in other circuits. This enables the main objective of achieving a simple recycling process that is eco-friendly through the use of IPA, which has negligible toxicity when at low concentrations.

To enable reuse of aggregates of Ag flakes formed during the recycling process, the aggregates were soaked in clean IPA, left to precipitate, and decanted four times to allow most of the PU residues to be separated from the silver. After the fourth wash, the IPA is decanted, and the silver is dried at room temperature. At this point, large pieces of PU residue (still with trace amounts of Ag), as seen in **Figure S10**, can be removed with a pair of tweezers. The recovered Ag-rich powder (**Figure S11**) can then be used to make a new ink by mixing them in pristine WPU dispersion following the initial method and quantities. The full process for washing the Ag flakes and recycling them in a new ink is detailed in the Methods Section. The downside of this method for washing the Ag Flakes multiple times in IPA is that, at the end, around 19.83 % of the initial ink is lost in the process. As described in the Methods Section, from an initial chunk of ink weighting 12g (containing 10.701 g of pristine Ag flakes and 1.298 g of polymer), we were able to recover 8.89 g of Silver powder with PU residues (which was used in the recycled ink), as well as 0.73 g of large PU chunks with silver residues that were discarded. More efficient separation processes, such as electrowinning[54] could also be employed for recovery of the rest of the Ag, at the cost of higher toxicity and increased complexity.

After printing the recycled ink, we observed changes in the morphology of the stencil printed traces. As shown in **Figure S12**, some PU residue from the initial ink are "trapped" inside the newly synthesized polymer, leading to a non-uniform rough surface, compared to the smooth surface of the initial printed traces. Nevertheless, the traces printed with recycled ink presented a conductivity of $1.13 \times 10^5$ S/m, which represents a decay of only ~2.4% compared to the conductivity of the pristine ink ($1.16 \times 10^5$ S/m, as shown in the previous section).

In terms of electromechanical properties, as shown in Figure S13, the recycled ink shows reduced stretchability when compared to the ink made of pristine silver. While the initial Ag-WPU tracks can withstand more than 25% strain without impacting the conductivity, in the recycled counterpart the resistance is shown to increase rapidly above 20% strains, leading to loss of functionality. While the issue of reduced stretchability of the recycled ink could be solved through the inclusion of LM, this would come at the cost of less straightforward and eco-friendly recycling. In applications that don't require stretchability but only flexibility (for instance smart labels as shown later), the benefits of scalable low-cost fabrication, and straightforward eco-friendly recycling process surpass the disadvantage of reduced stretchability, given that high electrical conductivity is maintained.

**Figure S14** compares the strain-stress curves of the pristine TPU film that is used as a substrate, with the TPU film that is recovered after undergoing the IPA-based circuit-degradation method. As can be observed, the proposed method for separation of the circuit's components doesn't impact the mechanical properties of the TPU film, which can be directly reused as a substrate for new prints after being dried. This happens since the IPA only leads to swelling of the TPU film, which is reversed after all IPA is evaporated.

2.4. Liquid metal inclusion

To overcome the electrical failure of the Ag-WPU ink at low strains (<30 %), the inclusion of eutectic gallium-indium (EGaIn) in the conductive compound was tested, aiming at creating a biphasic structure which, ideally, could deform without impacting its conductivity (Figure 5A). Details of the synthesis of the Ag-EGaIn-WPU biphasic mixture are presented in the Methods section. As observed in Figure 5B, the normalized resistance ($R/R_0$) for three Ag-EGaIn-WPU printed traces remains below 2 Ω/Ω when the sample is stretched up to ~20 %, for two of the samples, while the third sample shows a normalized resistance below 2 Ω/Ω up to ~24 % strain. At larger strains, the resistance increases for all three samples up until electrical failure at 203 %, 222% and 323.5 % strain, respectively. This failure was defined as occurring when the normalized resistance increased up to 100 Ω/Ω. The presence of EGaIn in the ink allows the samples to stretch ~10 times more before electrical failure when compared to the ink without liquid metal. Similarly, the electrical conductivity, reported in Figure S15, decreases slowly at a constant rate up to 200% strain from ~$10^5$ S/m down to $10^4$ S/m for 2 of the samples, while for a 3rd sample the sample conductivity decrease occurs only at ~400% strain. After these limits, the conductivity starts do decrease rapidly until complete electrical failure is observed. Interestingly, as depicted in **Figure S2**, the integration of EGaIn in the Ag-WPU ink doesn't lead to a noticeable change in the bulk initial conductivity of the ink which remains stable between 0 and 75% EGaIn wt%. Instead, the role of EGaIn in the ink is to create conductive paths that keep the percolative network created by Ag flakes stable at larger strains that when no EGaIn is used.

*In terms of the mechanical properties of the Ag-EGaIn-WPU, the full stress-strain curves for the three tested samples are shown in **Figure 5C**. As shown in **Figure 5D**, the mechanical failure of the samples occurred between 454.3 and 483.7% strain, similarly to what occurs in the samples without liquid metal. The elastic modulus of the Ag-EGaIn-WPU ink (**Figure 5E**) was estimated to be in the 13.79-15.9 MPa range, which is*

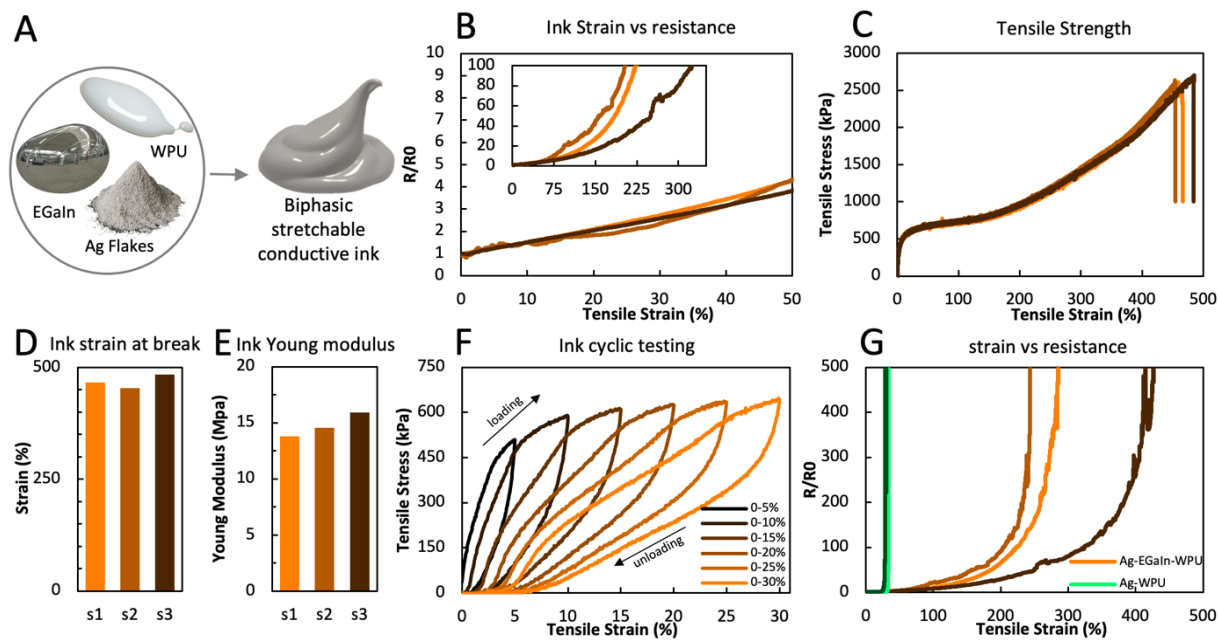

**Figure 5. A:** Inclusion of liquid metal (EGaIn) in the Ag-WPU ink leads to a biphasic stretchable conductive compound. **B:** Strain vs resistance curve for Ag-EGaIn-WPU traces printed over TPU substrate (three samples from distinct ink batches). **C:** Strain vs stress plot for the three printed ink samples. **D:** strain at break for the three ink samples. **E:** Estimated Young modulus for the 3 ink samples. **F:** Cyclic test of the Ag-WPU ink. **G:** Strains vs resistance curves for both Ag-WPU and Ag-EGaIn-WPU

*slightly higher than when EGaIn is not present. No smearing of EGaIn from the polymeric compound was observed during the tests.*

In Figure 5F, the cyclic performance of the Ag-EGaIn-WPU composite is shown. The sample was first stretched by 5% strain and then relaxed to its original length, and the strain was increased by 5% up to 30% for each successive cycle. While the maximum stress for each cycle follows similar values as for the strain-stress curve, a hysteresis loop between loading and unloading cycles can be observed. After 20% strain, plastic deformation appears to occur, leading to a permanent stretch of ~3.5% in the sample for the following cycles, similarly to the ink without EGaIn.

Figure 5G compares the strain-resistance curve for both polymers (with and without EGaIn). From the plots the resistance of the printed Ag-EGaIn-WPU traces increases much more slowly with stretch compared to Ag-WPU traces. Moreover, the biphasic Ag-EGaIn-WPU mixture can support much greater strains (i.e. >200%) before losing conductivity. However, these advantages in material performance are at the expense of greater complexity with recycling EGaIn-filled composites. When EGaIn is present in the conductive ink, the proposed IPA-based separation process no longer works since EGaIn alloys with the Ag flakes. In this case, we must modify our recycling process to incorporate the complex EGaIn separation processes described in [55].

In order to compare the cyclic stability of the inks, specimens of Ag-WPU and Ag-EGaIn-WPU were subjected to repetitive loading up to a strain of 10%, as depicted in **Figure S16**. The Ag-WPU sample exhibited an escalating trend in relative resistance with each successive loading cycle, resulting in a 23-fold increase in the relative resistance within the 50 cycles. Conversely, the sample incorporating liquid metal demonstrated enhanced stability to cyclic loading during the examination, displaying merely a 1.5-fold rise in its relative resistance after 50 loading cycles.

Despite the mechanical hysteresis observed in the samples, particularly in Figures 3G and 5F for the Ag-WPU and Ag-EGaIn-WPU materials, it is important to note that the impact of hysteresis on the electrical performance of thin film electronics can vary depending on the specific application. In the case of the Ag-WPU material, which exhibits lower stretchability and cyclic stability, the observed hysteresis may not have a significant impact on its electrical performance in certain applications. This ink formulation, although less stretchable, offers a level of flexibility that can be advantageous in certain scenarios. For instance, in the context of smart labels integrated into thin plastic wraps (as shown later), where moderate flexibility is required, the Ag-WPU material can be suitable. The hysteresis in this case is unlikely to hinder the functionality of the electronics.

On the other hand, if the application demands stretchability and higher stability under cyclic loading, the ink formulation containing EGaIn becomes more appropriate. The presence of EGaIn in the Ag-EGaIn-WPU ink enhances the stretchability and mechanical stability of the resulting thin film. This ink formulation demonstrates lower hysteresis, enabling better performance under repeated stretching and cycling.

Lastly, we observe that the microstructure of the Ag-EGaIn-WPU ink contains agglomerates of liquid metal that are fully encapsulated in PU, leading to a relatively rough surface for the printed traces compared to Ag-WPU (**Figure S17A**). To prevent nozzle clogging by the EGaIn agglomerates, the smallest usable nozzle for direct ink writing with Ag-EGaIn-WPU composite (**Figure S17B**) is 250um in diameter. Further research is necessary to fully understand the processes behind such EGaIn-induced agglomeration.

SEM and EDS imaging of Ag-EGaIn-WPU ink (**Figures S18 and S19**) shows a clear separation of Ag and Ga, while In is present all over the sample. This can be explained by the high affinity between Ag and In. Ag bonds to In, which is pulled out of EGaIn towards Ag particles, helping to anchor the EGaIn particles.

Moreover, in **Figures S20 and S21**, the formation of intermetallic Ag-In microparticles can be observed. These correspond to $AgIn_2$, which is formed at temperatures below 100 ºC and, while having the same size of Ag flakes (< 5 μm), Ag-In particles have a distinctive spherical shape.

2.5. Applications

The Ag-WPU materials architecture and fabrication techniques enable the rapid implementation of soft circuits with integrated microchips and SMD components that can be easily recycled. Possible application domains for this system include recyclable product packaging and on-skin electronics for health monitoring, which rely on highly flexible, skin-conformal circuits that must be comfortable to wear. As a first use-case, we present a fully recyclable smart package that continuously monitors the temperature of perishable products during handling and provides information to consumers regarding any mishandling or previous storage in non-ideal conditions. For the second use-case, we present a family of physiological sensing stickers based on Ag-WPU ink and integrated microchips for the acquisition of multiple digital biomarkers, including continuous axillary temperature monitoring, single-lead electrocardiography (ECG) monitoring, and surface electromyography (sEMG) recording for hand gesture and facial expression recognition. In these demonstrations the Ag-WPU ink formulation (without liquid metal) was used to enable fully recyclable systems.

*2.5.1. Fully recyclable smart packaging*

Foodborne illness caused by ingestion of contaminated food can pose a serious public health threat. One of the most prominent causes for contamination is related to poor refrigeration of fresh and perishable products since bacteria can multiply rapidly when products are left above 5 ºC for extended periods [56]. A printed smart label is introduced to monitor the storage temperature of perishable products. The recyclable label can be integrated into various packaging forms, such as a recyclable water bottle (Figure 6A) or a fresh fish package (Figure 6B), to help merchants and consumers identify poor handling rapidly. As shown in Figure 6B for a package of fresh salmon, a green LED is lit up whenever the perishable good has been properly stored at temperatures below 5 ºC. If the temperature rises above 5 ºC for longer than one hour, a red LED lights up and stays on even if the package is later brought back to a safe storage temperature, as shown in the temperature plots from **Figure S22**. The BLE connection to the smart tag allows the user to review the historical temperature data as well as the current storage temperature of the package. The functioning of the smart label system is shown in Videos S3 and S4.

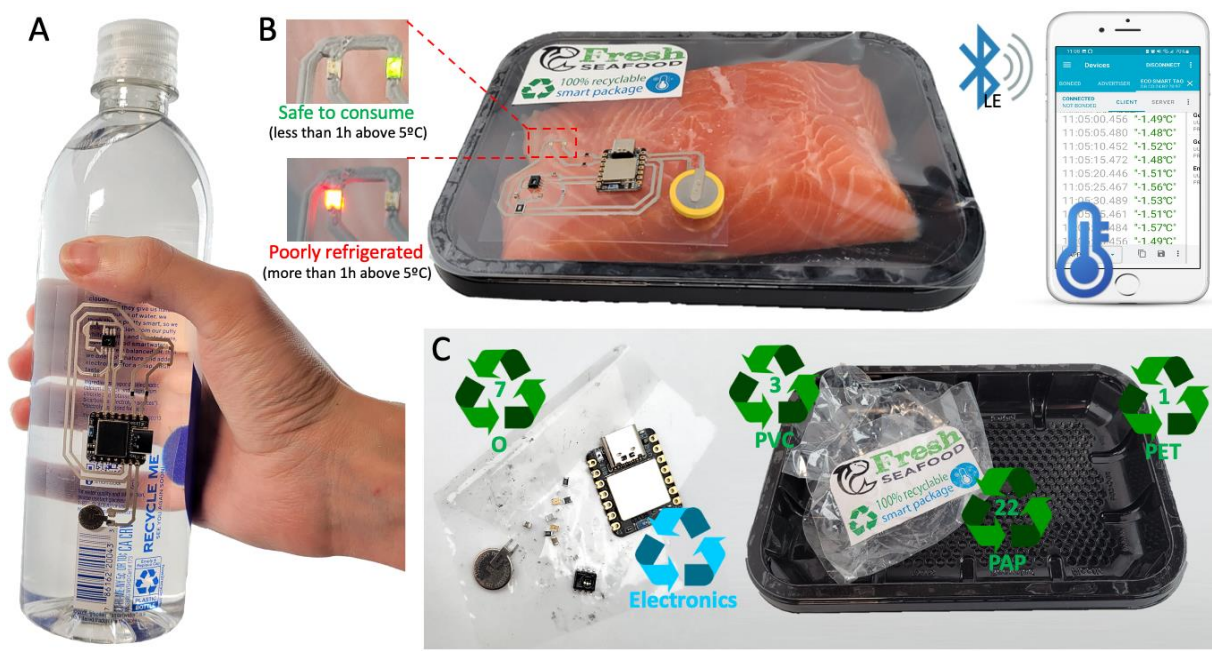

**Figure 6.** Fully recyclable smart packages. **A:** Printed recyclable smart label integrated in a plastic water bottle. **B:** A printed smart label integrated in a package of perishable fresh food monitors its storage temperature over time and informs consumer through two LEDs whether the item is safe to consume or whether there is the risk of contamination due to poor refrigeration. Historic and current temperature data can be transmitted via BLE to a smartphone. **C:** Both the food package and the integrated smart label can be fully recycled using conventional means and the proposed separation and recycling method, respectively. Materials that can be recycled through conventional methods are identified by their United Sates recycling codes (PET – 1; PVC – 3; other plastics – 7; paper - 22).

As shown in Figure 6C, the full package and smart label can be recycled. The plastic package is composed of PET, PVC, and paper that can be recycled in conventional ways (as identified by the United States recycling codes in the figure), while the smart label can be recycled through our proposed method for recycling the printed electronic circuits.

*2.5.2. Body temperature monitoring wearable*

A microcontroller-based system with WiFi capabilities was designed to measure body temperature through a thermistor. The circuit shown in **Figure 7A** was digitally printed through direct ink writing over a soft TPU substrate using the Ag-WPU conductive polymer and it was then transferred to a double-side medical-grade adhesive film. All rigid SMD components were bonded to the printed lines through the proposed soft soldering process (**Figure 7B**) and the outline of the patch was laser-cut.

After calibrating the system, the patch was transferred to the user's chest, with the measuring thermistor placed directly on the user's axilla, as shown in **Figure 7C**. The sensor was worn for 1 hour while the user worked on a computer sitting at a desk, and it showed no signs of delamination or functionality issues. The temperature was measured by the patch at a frequency of 0.2 Hz. As ground truth, the temperature was also measured using a commercial digital thermometer. Figure 7D shows the acquired signals, indicating that the patch temperature follows the ground truth measurement trend. The average temperature for both the patch and ground truth is 36.2 ºC.

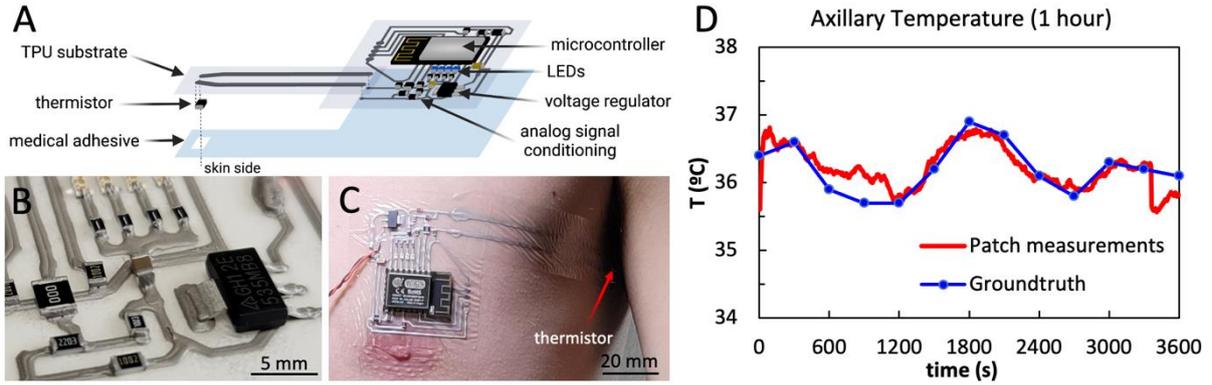

**Figure 7. A:** Design of multilayer printed patch which includes a microcontroller, LEDs, voltage regulator and analog signal conditioning circuitry for thermal sensing through a skin-contact thermistor. **B:** Detail of the printed circuit where the interface between rigid SMD components and the soft printed circuit can be observed. **C:** Thermal sensing patch adhered to the skin, with the thermistor in the armpit region. **D:** Axillary temperature measured in the armpit for 1 hour using the patch (in red, data acquired every 5 s, filtered with a moving average filter with a 60 second window) and a ground truth obtained using a commercial digital thermometer (in blue, data acquired every 5 s).

*2.5.3. Multielectrode patch for electrophysiology*

To record biopotentials from the body, a soft patch with three conductive electrodes was printed on a TPU substrate, and a SMD connector and medical adhesive were added to the patch to ensure good contact with the skin, as detailed in the Methods section. The patch's structure is displayed in Figure 8A.

The multielectrode patch was placed over the flexor carpi radialis muscle in a volunteer's forearm to acquire electromyography (EMG) signals, as shown in Figure 8B. The Ag-WPU was also used to detect surface EMG activity produced by a finger performing flexion or extension, as shown in Figure 8C. When gripping a handheld dynamometer (Figure 8D), the wrist flexors all contract simultaneously and generate EMG signals, as shown in the plots from Figure 8E, with the amplitude growing proportionally with the applied gripping force.

The same bioelectronic patch can also be placed on the forehead near the temple, as shown in Figure 8F. In this case, the task consisted of performing twice the same sequence of facial expressions: smiling, jaw clenching, eye blinking, and eyebrow flashing, before returning to a resting state. In Figure 8G, the top plot shows evidence of the repeatability of the signal morphology for each facial expression across the two repetitions. Likewise, the bottom plot corresponding to the amplitude RMS envelope of the above signal shows that both the amplitude and signal shape is the same for the repetitions and distinct enough among different actions.

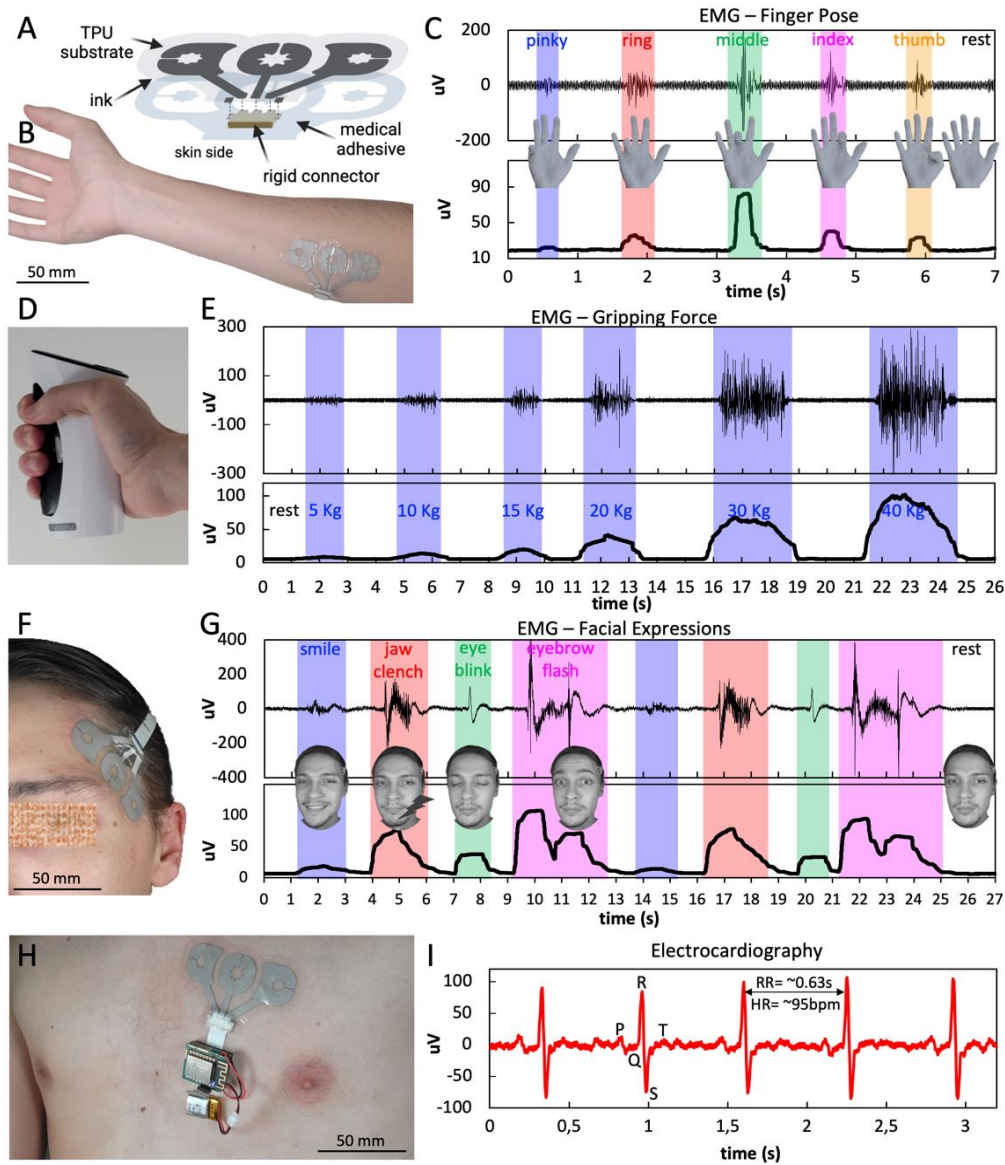

**Figure 8. A:** Printed multi-electrode patch with integrated flat-cable connector for skin-surface electrophysiology. **B:** Electrophysiology patch adhered to the right forearm over the flexor carpi radialis muscle. **C:** EMG signals and corresponding RMS amplitudes produced by the flexion/extension of different fingers. **D:** Hand dynamometer used to measure hand grip strength. **E:** EMG signal and corresponding RMS amplitude envelope acquired during a task consisting of squeezing the dynamometer with increasing gripping forces, with rests in between. **F:** Multi-electrode patch adhered on the face near the temple. **G:** Bioelectronic signals produced by various facial expressions and subtle movements, including smiling, jaw clenching, eye blinking, and eyebrow flashing. Bottom plot shows the corresponding RMS amplitude envelope for each expression. **H:** Electrophysiology patch adhered to the user's chest, over the sternal portion of the left pectoralis major muscle, connected to a biopotential recording system and battery. **I:** ECG signal acquired with the printed patch, where ECG features (P and T waves, QRS complex) can be observed. RR interval and heart rate (HR) can also be calculated.

In the above facial gesture case, the various actions performed in each task are quite distinct from one another and similar between repetitions, as evidenced by the Dynamic Time Warping plots in **Figures S23-S26**. **Figure S27** shows the Euclidean distance between pairs of facial EMG signals for the 2 repetitions of the task, which correlates to the similarity between pairs of signals. In this figure, the lowest distance between two gestures of distinct repetitions is highlighted, and accurately corresponds to the same gesture in the two repetitions for the four tested cases. In this sense, a simple classifier based on dynamic time warping could be reliably employed for basic task classification and their eventual application in human-machine interfaces [57–59]. Moreover, the presented use cases provide a measurable way to assess everyday motor tasks, serving as digital biomarkers of movement. Clinicians can adopt these biomarkers to monitor the progression of neuromuscular disorders and other clinical conditions with motor symptoms. [60–63].

In Figure 8H, the same electrode patch as before was adhered to the chest over the sternal portion of the left pectoralis major muscle and used to acquire electrocardiography (ECG) signal, which is shown in Figure 8I. In the depicted signal, the standard features of a normal ECG wave can be observed: the QRS complex, as well as P and T waves, which are labeled in the figure. The RR interval (corresponding to the time between two consecutive R peaks) was calculated as being 630 ms, from which a heart rate (HR) of 95 bpm can be estimated. This HR value is within the normal limits stipulated by physicians for a healthy heart [64].

As shown in **Figure S28**, the electrode-skin impedance of the printed Ag-WPU electrodes was measured and compared to that of conventional Ag/AgCl electrodes, which are commonly used in healthcare practice. One of the conclusions that can be drawn is that the printed electrodes show a lower intrinsic electrical resistance than the Ag/AgCl material, evidenced by their lower impedance at high frequencies (~100 kHz). Nevertheless, Ag/AgCl electrodes present lower interface resistance with the body, supported by the lower impedance at low frequencies (20 kΩ at 1 Hz, compared to 400 kΩ of the Ag-WPU electrodes at the same frequency). Moreover, the overall lower impedance of Ag/AgCl electrodes in the tested frequency range (1 Hz to 100 kHz) indicates that they allow for a low interface capacitance. This can be explained by the presence of a wet hydrogel, which creates a soft ion-rich medium that conforms to the skin and improves charge flow between the body and the recording electronics. This contrasts with what happens with dry printed Ag-WPU electrodes. Nonetheless, despite their slightly higher impedance (which may be translated to noisier measurements), the dry thin-film printed electrodes show some advantages related to the straightforward fabrication of user and application-specific biopotential recording patches with custom number and positioning of electrodes as well as intricate electrode shapes. Moreover, the fact that these printed electrodes are dry (i.e. no wet gel or electrolyte-rich conductive paste is interfacing them with the skin) decreases the chances of electrode cross-talk or electrode-skin interface degradation over time due to electrolyte drying, as previously discussed in Ref. [65].

3. Conclusion

In this study, we introduce a novel eco-friendly class of soft conductive inks composed of combinations of water-based polyurethane, silver, and liquid metal that exhibit high electrical conductivity (1.1 - 1.6x10$^5$ S/m), is suitable for digital printing methods like DIW, and can be printed on thin-film substrates to create highly flexible, high resolution (down to 200 μm line width), sticker-like electronics. Contrary to previous works reporting printable conductive mixtures, the Ag-WPU ink remains stable for up to 4 weeks at room temperature without affecting conductivity or printability. By taking advantage of the robust adhesive properties of the proposed conductive polymer, a simple "soft soldering" process is proposed allowing for direct integration of IC chips and miniaturized SMD components into the printed circuits without the need of other electrically conductive adhesives or complex bonding processes. Moreover, the entire process can be performed at room temperature without the need for sintering, resulting in low-cost and fast fabrication of functional chip-integrated circuits. Electrical failure occurs at ~30% strains, but the strong mechanical bonding between the printed Ag-WPU traces and SMD components allows for circuits to stretch up to 380% before mechanical failure.

Another key feature is that the Ag-WPU ink is fully recyclable and compatible with environmentally sustainable practices: The ink is dispersed in water and free of organic solvents, and the chip-integrated soft circuits can be disassembled by soaking and stirring in isopropyl alcohol, leading to the separation of the rigid components, TPU substrate film, and ink residues. At this stage, they can be properly discarded, reprocessed (in the case of the ink), or directly reused (in the case of the SMD components and substrate). To complete the circular life cycle, the separated ink aggregates can be washed multiple times in IPA to separate Ag flakes and PU residues. The recovered silver is then used to synthesize a new ink that exhibits 97.6% of the initial ink's conductivity. This straightforward recycling method can be implemented without the need for complex equipment or harmful chemicals, contrasting with the few degradation and recycling processes proposed for soft electronics that are complex, labor intensive and usually lead to environmentally harmful chemical byproducts.

In order to make the ink more mechanically robust, EGaIn was added to form a biphasic composite. This led to a conductive ink that can be stretched to more than 200% strain without electrical failure. This added stretchability with EGaIn, however, is at the expense of a more complex decomposition and recycling method, which is needed to allow full separation of the metals.

We showcase the potential of this material system by creating on-skin bioelectronic patches that acquire multiple digital biomarkers, including body temperature, and electrophysiological signals related to heart activity and muscle activity during motor tasks. Finally, we present a smart package in which Ag-WPA circuitry is used to create an electronic printed label that records the storage temperature of perishable items and warns consumers about unsafe storage conditions that can lead to product contamination. The full smart package is recyclable by employing both conventional methods (for the food packaging) and the proposed separation and recycling method for the soft electronic temperature monitor. Together, these implementations demonstrate the potential for Ag-WPA inks to support electronic sticker functionality while reducing material waste.

Summing up, we demonstrate for the first time a conjunction of simple materials and methods that enable complex eco-friendly soft electronic circuits with a fully circular life-cycle. These materials and methods include the development of an eco-friendly conductive polymer compatible with digital printing methods, a facile method for room-temperature, sinter-free, reliable integration of SMD chips using the same conductive polymer, and a method for straightforward and low cost separation and degradation of the fabricated circuits that enables the recovery and reuse of the old circuit's constituents for use in new circuits across various fields.

4. Methods

4.1. Conductive ink preparation

First, 1.25 g of an organic-solvent-free waterborne polyurethane (WPU) dispersion (U4101 40% solid content, Alberdingk) are transferred to a 20 ml capacity glass vial. Next, 4.12 g of silver flakes (Silflake 071, Technic inc.) are mixed into the WPU solution (3.296:1 Ag:WPU wt.%) using a planetary mixer (Thinky AR-100) for 3 minutes at 2000 rpm. After mixing in the Ag flakes, a thick paste is obtained. It is diluted by adding 0.5 g of deionized water and the compound is again mixed in the planetary mixer for 3 minutes at 2000 rpm. The presented quantities lead to 89.18% concentration of Ag flakes in the final ink after all water is evaporated.

For the ink samples including liquid metal (Ag-EGaIn-WPU), we first synthesized the Ag-WPU ink as described above. After this, 6.2 g of EGaIn (75.5% Ga, 24.5% In) were heated to 60 ºC and added to the Ag-WPU ink. After the inclusion of EGaIn, the polymer was mixed for 10 minutes at 500 rpm using a vertical overhead stirrer instead of the planetary mixer, since the vertical mixer reduces the size of EGaIn-PU agglomerates due to the higher shear, thus preventing nozzle clogging. The process is shown in **Figure S29**.

4.2. Stencil printing

Using a $CO_2$ desktop laser cutting system (Universal Laser Systems VLS3.50), the shapes to be printed are patterned on the stencil material (Blazer Orange Laser Mask, Ikonics Imaging). The stencil is then adhered to the desired substrate and ink is spread using a single edge razor blade. The stencil is immediately lifted and removed, and the print is let to dry at room temperature for 5 minutes before further processing steps. The used stencil material leads to printed ink layers with 102 μm height.

4.3. Direct ink writing

The Ag-WPU solution is loaded into a syringe barrel for direct ink write (DIW) printing. The ink is then dispensed over a TPU film (Bemis 3412 TPU hotmelt film) following defined circuit paths using a Voltera ink dispensing system (Voltera V-One PCB printer). Next, the film is left to dry at room temperature for 5 minutes before integrating the microchips and SMD components. The direct ink writing process is show in **Video S5**.

When printing with Ag-EGaIn-WPU, the nozzle diameter limit is 250um since at lower gauges nozzles will clog due to LM-PU agglomerates.

**4.4. Conductivity measurement and aging tests**

Each sample in the conductivity test consists of five stencil-printed traces (dimensions 8 cm x 5 mm x 102 μm) printed over an FR1 rigid substrate (FR1 Substrate 3" x 4", Voltera), as shown in **Figure S30**. At the each end of each trace, a droplet of EGaIn was dispensed to reduce the contact resistance between the printed polymer and the multimeter probes. The resistance of each sample is measured by a digital desktop multimeter (2100, Keithley) with a four-point probe. For each track, 3 measurements were taken.

For the aging test of the printed lines, the print was stored uncovered at room temperature. For the storage test, various ink batches were fabricated and stored in closed glass vials at room temperature before printing the samples over time. Each day, a new set of five traces was printed using the aged ink vial after mixing the ink for 3 minutes at 2000 rpm, and its resistance was measured after letting the print dry.

**4.5. Tensile testing**

TPU samples were patterned into a standardized dogbone shape (Die C, ASTM D412) by cutting with a $CO_2$ laser cutter (VLS3.50; Universal Laser System). For characterization of the Ag-WPU ink, an ink trace was printed on each dogbone using a stencil, as previously described. Dimensions are shown in **Figure S31**. To analyze the electromechanical failure modes of the "soft solder" joints, two conductive traces were printed on each dogbone, and 0 Ω SMD resistors (1210 and 0805 sizes) were bonded to the traces, as depicted in **Figure S32**.

Tensile testing was done on a materials testing machine (5969; Instron) with a 10 N load cell to which the samples were clamped. For the quasi-static strain-stress characterization and cyclic testing, the strain loading rate was set to ~0.1%/s until the sample broke.

To measure sample resistance, a digital desktop multimeter (2100, Keithley) with a four-point probe was connected to a laptop via USB and measurements were acquired using KickStart Instrument Control Software (Version 2.9.0. for windows 10).

To estimate conductivity of the ink samples upon stretching, it was assumed that the conductive materials were incompressible, maintaining a constant volume during deformation. The volume ($V$) of the conductive sample is given by **Equation 1** where $l$ corresponds to the sample's length, $t$ to its thickness, and $w$ to its width:

$$V = l \times t \times w \Leftrightarrow t \times w = V/l \tag{1}$$

The conductivity, assuming constant volume, is then given by Equation 2:

$$\sigma = \frac{l}{R \times w \times t} \Leftrightarrow \sigma = \frac{l}{R \times \left(\frac{V}{l}\right)} \Leftrightarrow \sigma = \frac{l^2}{R \times V} \tag{2}$$

It's important to note that the assumption of incompressibility of the conductive materials results in only an approximate estimate of the sample's conductivity upon stretching.

4.6. Separation and recycling of Silver from recovered ink chunks

To separate the Ag flakes from the PU residues, the recovered ink aggregates were soaked in clean IPA. 5 g of clean IPA are used for each 1g of recovered ink. After manually stirring the contents, these are left to precipitate for 1 hour, and the IPA is decanted. The process is repeated four times to allow most of the PU residues to be separated from the silver. After the fourth wash, the IPA is decanted, and the solid contents are dried at room temperature and scraped for the wall and bottom of the container using a spatula. The larger pieces of cured PU (still with traces of Ag) are removed with tweezers and discarded. The remaining powder, consisting of recovered Ag flakes and PU residues can then be used to make a new ink by mixing 1.25 g of pristine WPU with 4.12 g of recovered silver powder and 0.5 g of DI water).

To test the conductivity of the recycled ink, 12g of cured ink pieces (1.298 g PU, 10.701 g Ag) were subjected to the recycling process. The weight of solid contents after the fourth IPA wash was 9.62 g (80.16 % of the initial weight), from which 0.73 g of large PU chunks were discarded and 9.89 g of Ag flakes-rich powder were recovered.

After synthesizing a new ink with the recovered Ag-rich powder, five traces were stencil-printed and their conductivity measured, as described above in Sec. 4.4. Printed traces of recycled ink are shown in **Figure S33**.

4.7. Smart packaging

A circuit containing an $I^2C$ digital humidity/temperature sensor (HIH6139, Honeywell) and a Bluetooth low energy (BLE) enabled microcontroller (Seeed Xiao nRF52840, Seeed Studio) was designed and printed using the proposed Ag-WPU conductive ink, as shown in **Figure S34**. This circuit also contains a 3.7 V rechargeable LiPo coin cell battery, SMD resistors and capacitors, and two LEDs to provide information to the user.

To acquire temperature data over BLE, a debug app (nRF Connect, Nordic Semiconductors) was installed in a smartphone (S21 5G, Samsung). Temperature and humidity data was transferred to the mobile phone every time the app was paired with the printed smart tag.

For the fully recyclable smart package demo, a polyethylene terephthalate (PET) deli container was used. The transparent plastic cover was made of polyvinyl chloride (PVC) plastic wrap.

4.8. On-skin temperature monitor

A circuit containing a WiFi-enabled microcontroller (ESP8266), NTC thermistor (10 kΩ at 25ºC, 1206) and voltage divider for signal conditioning, as well as 4 LEDs was designed and printed in a TPU substrate (Bemis 3412 TPU hotmelt) through direct ink writing. After printing, the flexible circuit was transferred to a double side medical-grade adhesive (152A Medical Transfer Adhesive, 3M) and the SMD chips were bonded in place. The system was powered through an external 3.7V LiPo battery connected by copper wires.

The temperature sensing patch was calibrated using a hot plate (Cimarec, Thermo Scientific) with an attached digital thermometer (TP-16, ThermoPro), as shown in **Figure S35**. The patch was placed on the hotplate, with the thermistor near the digital thermometer. The ADC output read from the microcontroller was sent to a nearby computer via an ESP-Now wireless protocol. Simultaneously, the temperature of the hot plate was also recorded. The samples were tested between a range of 25 and 50 °C in 1 °C increments. We waited for 15 min between each data point recording to ensure that the temperature had reached equilibrium. 5 trials (3 with the temperature increasing and 2 with temperature decreasing) were performed and the calibration curve was then calculated. Due to the narrow temperature range, the sensor's response could be approximated linearly as shown in **Figure S36**.

The patch was adhered to the user's chest after carefully cleaning the skin with rubbing alcohol. Attention was given so that the thermistor would be placed directly on the user's axilla. Temperature was measured every 5 seconds for a total of 1 hour using the patch. Also, axillary temperature was manually measured every 5 minutes using a digital thermometer (KD-1761, 180 Innovations, 0.2ºC accuracy) in order to obtain a ground truth recording. The data acquired from the patch was filtered using a moving average filter (60 samples) and the values were sent in real time to a nearby computer via ESP-Now wireless protocol.

4.9. Biopotential recording electrodes

The three conductive electrodes were printed simultaneously by stencil printing over a TPU substrate film (Bemis 3412 TPU hotmelt) and a flat cable connector (Molex 52207-0933, 1mm pitch) was integrated in the patch using the described soft soldering method. The electrode patch was then transferred to a double side medical-grade adhesive (152A Medical Transfer Adhesive, 3M) that was previously patterned to shape using a $CO_2$ desktop laser cutting system (Universal Laser Systems VLS3.50) and finally, the outline of the patch was cut using the $CO_2$ laser.

The patch was connected to a miniaturized biopotential recording system previously presented in [65] through a flat cable and the electrophysiological recordings were sent by the analog front end (AFE) via UDP protocol and acquired by the OpenBCI GUI (v5.0.9 for Mac) and later converted to CSV files. The full acquisition setup is shown in **Figure S37**.

For recording of the biopotential signals, the user's skin was first carefully cleaned with rubbing alcohol and let to dry, before laminating the electrode patch and recording analog front-end into the recording site.

In this setup, acquired signals were sampled at 250 Hz and amplified 24×. They were then filtered in MATLAB using a notch filter (60 Hz) and High/Low pass second-order Butterworth filters as needed for each monitored signal. For ECG, the relevant frequencies were assumed to be in the 5–55 Hz band. For surface EMG (both in the arm and face), the relevant frequencies were in the 2–100 Hz band. For each acquired EMG signal, its corresponding root-mean square (RMS) envelope was as well calculated using a window with a length of 50 samples for the finger pose experiment and 250 samples for the gripping force and facial expressions experiments.

4.10. Electrode-skin impedance measurement

The electrode-skin impedance for the fabricated electrodes was measured using a PalmSens4 impedance analyzer. 50 impedance points (≈9.8 per decade) were measured between $10^0$ and $10^5$ Hz. Before placing the electrodes on the right inner forearm of the volunteer, the skin was cleaned by wiping with rubbing alcohol and dried for 1 min. The electrode patch was then placed and left to rest for 1 min, and the impedance measurement was taken. After the measurement was complete, the electrodes were removed, and any residue of adhesive was wiped off with rubbing alcohol.

Impedance of medical-standard Ag/AgCl electrodes (RedDot, 3M) was measured in the same frequency range by adhering three electrodes to the user's forearm after following the same skin cleaning procedure described above. Gripping force was measured through a handheld dynamometer (Digital Hand Dynamometer, Handeful).

The experiments in human subjects were approved by the Carnegie Mellon University Institutional Review Board (STUDY2022_00000015) in accordance with the US HHS regulations for the protection of human subjects in research at 45CFR 46. Informed consent was obtained from the volunteer, who is also the first author of the manuscript, and all experiments were performed in accordance with the applicable regulations.

4.11. SEM imaging

To analyze morphologic changes in Ag flakes, samples of pristine and recovered Ag flakes were prepared and imaged using a FEI Quanta 600 FEG scanning electron microscope with an acceleration voltage of 15.00 kV and magnification of 5000x. Images were taken on carbon adhesive disks and the scale bar corresponds to 10 μm.

The Ag-EGaIn-WPU samples were immersed for 90 s in liquid nitrogen and fractured through a mechanical impact, leading to a clean cross-section fracture. EDS surface scanning was used to build the color map of the distribution of the elements (Ga, Ag, and In) on the sample's surface.

4.12. Rheological characterization

Rheological measurements were performed on a DHR-2 stress-controlled rheometer (TA Instruments) at 25 ºC for Ag-WPU samples that had been stored at room temperature for 0 to 35 days. A 40-mm parallel plate configuration was used with a 600 µm gap height.

4.13. Statistical analysis

Unless otherwise stated, one measurement per sample was taken. Data treatment was performed in Matlab_R2020b and Microsoft Excel V16.65 for Mac.

Acknowledgments

Support for this research was provided by the Fundação para a Ciência e a Tecnologia (Portuguese Foundation for Science and Technology) through the Carnegie Mellon Portugal Program under Grant SFRH/BD/150691/2020. Support also came from the European Commission through the European Research Council project Liquid3D | GA 101045072 | ERC-2021-COG and from the CMU-Portugal project WoW (45913), which had the support of the European Regional Development Fund (ERDF) and the Portuguese State through Portugal 2020 and COMPETE 2020.

**References**

1. Kumar, A., Holuszko, M. & Espinosa, D. C. R. E-waste: An overview on generation, collection, legislation and recycling practices. *Resources, Conservation and Recycling* vol. 122 (2017).
2. Grant, K. *et al.* Health consequences of exposure to e-waste: A systematic review. *Lancet Glob. Heal.* **1**, (2013).
3. Tavakoli, M. *et al.* EGaIn-Assisted Room-Temperature Sintering of Silver Nanoparticles for Stretchable, Inkjet-Printed, Thin-Film Electronics. *Adv. Mater.* **30**, (2018).
4. Joshipura, I. D., Ayers, H. R., Majidi, C. & Dickey, M. D. Methods to pattern liquid metals. *Journal of Materials Chemistry C* vol. 3 3834–3841 (2015).
5. Kramer, R. K., Majidi, C. & Wood, R. J. Masked deposition of gallium-indium alloys for liquid-embedded elastomer conductors. *Adv. Funct. Mater.* **23**, 5292–5296 (2013).
6. Alberto, J. *et al.* Fully Untethered Battery-free Biomonitoring Electronic Tattoo with Wireless Energy Harvesting. *Sci. Rep.* (2020) doi:10.1038/s41598-020-62097-6.

7. Costa, G. *et al.* 3D Printed Stretchable Liquid Gallium Battery. *Adv. Funct. Mater.* (2022) doi:10.1002/adfm.202113232.

8. Carneiro, M. R. & Tavakoli, M. Wearable Pressure Mapping through Piezoresistive C-PU Foam and Tailor-Made Stretchable e-Textile. *IEEE Sens. J.* **21**, (2021).

9. Kang, S. *et al.* Highly Sensitive Pressure Sensor Based on Bioinspired Porous Structure for Real-Time Tactile Sensing. *Adv. Electron. Mater.* **2**, (2016).

10. Liu, S., Shah, D. S. & Kramer-Bottiglio, R. Highly stretchable multilayer electronic circuits using biphasic gallium-indium. *Nat. Mater.* **20**, (2021).

11. Kim, D.-H. *et al. Epidermal Electronics*. http://science.sciencemag.org/.

12. Libanori, A., Chen, G., Zhao, X., Zhou, Y. & Chen, J. Smart textiles for personalized healthcare. *Nature Electronics* vol. 5 (2022).

13. Neumann, T. V. & Dickey, M. D. Liquid Metal Direct Write and 3D Printing: A Review. *Advanced Materials Technologies* vol. 5 (2020).

14. Baharfar, M. & Kalantar-Zadeh, K. Emerging Role of Liquid Metals in Sensing. *ACS Sensors* **7**, 386–408 (2022).

15. Yang, J., Cheng, W. & Kalantar-Zadeh, K. Electronic Skins Based on Liquid Metals. *Proceedings of the IEEE* vol. 107 (2019).

16. Wu, W. Inorganic nanomaterials for printed electronics: A review. *Nanoscale* **9**, 7342–7372 (2017).

17. Tian, B. *et al.* All-printed, low-cost, tunable sensing range strain sensors based on Ag nanodendrite conductive inks for wearable electronics. *J. Mater. Chem. C* **7**, 809–818 (2019).

18. Tian, B. *et al.* Fully Printed Stretchable and Multifunctional E-Textiles for Aesthetic Wearable Electronic Systems. *Small* **18**, 1–14 (2022).

19. Wiklund, J. *et al.* A review on printed electronics: Fabrication methods, inks, substrates, applications and environmental impacts. *Journal of Manufacturing and Materials Processing* vol. 5 (2021).

20. Ghosh, B., Ghosh, M. K., Parhi, P., Mukherjee, P. S. & Mishra, B. K. Waste Printed Circuit Boards recycling: An extensive assessment of current status. *Journal of Cleaner Production* vol. 94 (2015).

21. Moin, A. *et al.* A wearable biosensing system with in-sensor adaptive machine learning for hand gesture recognition. *Nat. Electron.* **4**, (2021).

22. Tan, P. *et al.* Solution-processable, soft, self-adhesive, and conductive polymer composites for soft electronics. *Nat. Commun.* **13**, (2022).

23. Held, M. *et al.* Soft Electronic Platforms Combining Elastomeric Stretchability and Biodegradability. *Adv. Sustain. Syst.* **6**, (2022).

24. Lee, S. P. *et al.* Highly flexible, wearable, and disposable cardiac biosensors for remote and ambulatory monitoring. *npj Digit. Med.* **1**, (2018).

25. Lee, H. *et al.* Wearable/disposable sweat-based glucose monitoring device with multistage transdermal drug delivery module. *Sci. Adv.* **3**, (2017).

26. Fernandes, D. F., Majidi, C. & Tavakoli, M. Digitally printed stretchable electronics: A review. *Journal of Materials Chemistry C* vol. 7 14035–14068 (2019).

27. Ohm, Y. *et al.* Publisher Correction: An electrically conductive silver–polyacrylamide–alginate hydrogel composite for soft electronics. *Nat. Electron.* **4**, (2021).

28. Huang, Z. *et al.* Three-dimensional integrated stretchable electronics. *Nat. Electron.* **1**, (2018).

29. Saborio, M. G. *et al.* Liquid Metal Droplet and Graphene Co-Fillers for Electrically Conductive Flexible Composites. *Small* **16**, (2020).

30. Vaseem, M., McKerricher, G. & Shamim, A. Robust Design of a Particle-Free Silver-Organo-Complex Ink with High Conductivity and Inkjet Stability for Flexible Electronics. *ACS Appl. Mater. Interfaces* **8**, (2016).

31. Lopes, P. *et al.* Bi-Phasic Ag-In-Ga Embedded Elastomer inks for Digitally Printed, Ultra-Stretchable, Multi-Layer Electronics. *ACS Appl. Mater. Interfaces* **13**, 14552–14561 (2021).

32. Zu, W. *et al.* A Comparative Study of Silver Microflakes in Digitally Printable Liquid Metal Embedded Elastomer Inks for Stretchable Electronics. (2022) doi:10.1002/admt.202200534.

33. Li, J., Johnson, C. M., Buttay, C., Sabbah, W. & Azzopardi, S. Bonding strength of multiple SiC die attachment prepared by sintering of Ag nanoparticles. *J. Mater. Process. Technol.* **215**, (2015).

34. Chang, C. W., Cheng, T. Y. & Liao, Y. C. Encapsulated silver nanoparticles in water/oil emulsion for conductive inks. *J. Taiwan Inst. Chem. Eng.* **92**, (2018).

35. Ibrahim, N., Akindoyo, J. O. & Mariatti, M. Recent development in silver-based ink for flexible electronics. *Journal of Science: Advanced Materials and Devices* vol. 7 (2022).

36. Li, Z. *et al.* Highly Conductive, Flexible, Polyurethane-Based Adhesives for Flexible and Printed Electronics. *Adv. Funct. Mater.* **23**, 1459–1465 (2013).

37. Lall, P., Narangaparambil, J. & Miller, S. Development of Multi-Layer Circuitry Using Electrically Conductive Adhesive and Low-Temperature Solder Material for Surface-Mount Component Attachment. *Proc. ASME 2021 Int. Tech. Conf. Exhib. Packag. Integr. Electron. Photonic Microsystems, InterPACK 2021* (2021) doi:10.1115/IPACK2021-74086.

38. Lall, P., Goyal, K., Schulze, K. & Miller, S. Electrically Conductive Adhesive Interconnections on Additively Printed Substrates. *Intersoc. Conf. Therm. Thermomechanical Phenom. Electron. Syst. ITHERM* **2021**-**June**, 807–817 (2021).

39. Wu, H. *et al.* A highly conductive thermoplastic electrically conductive adhesive for flexible and low cost electronics. *Proc. Electron. Packag. Technol. Conf. EPTC* 1544–1546 (2014) doi:10.1109/ICEPT.2014.6922948.

40. Lee, J., Sjoberg, J., Rooney, D. T., Geiger, D. A. & Shangguan, D. Process development and reliability evaluation of electrically conductive adhesives (ECA) for low temperature SMT assembly. *Proc. IEEE/CPMT Int. Electron. Manuf. Technol. Symp.* (2008) doi:10.1109/IEMT.2008.5507852.

41. Sudheshwar, A., Malinverno, N., Hischier, R., Nowack, B. & Som, C. The need for design-for-recycling of paper-based printed electronics – a prospective comparison with printed circuit boards. *Resour. Conserv. Recycl.* **189**, 106757 (2023).

42. Jansson, E. *et al.* Suitability of Paper-Based Substrates for Printed Electronics. *Mater. 2022, Vol. 15, Page 957* **15**, 957 (2022).

43. Dulal, M. *et al.* Toward Sustainable Wearable Electronic Textiles. *ACS Nano* **16**, 19755–19788 (2022).

44. Arroyos, V. *et al.* A Tale of Two Mice: Sustainable Electronics Design and Prototyping. *Conf. Hum. Factors Comput. Syst. - Proc.* (2022) doi:10.1145/3491101.3519823.

45. Borland, H., Bhatti, Y. & Lindgreen, A. Sustainability and sustainable development strategies in the U.K. plastic electronics industry. *Corp. Soc. Responsib. Environ. Manag.* **26**, 805–818 (2019).

46. Zhou, Y., Xu, Z., Bai, H. & Knapp, C. E. Room Temperature Electronic Functionalization of Thermally Sensitive Substrates by Inkjet Printing of a Reactive Silver-Based MOD Ink. *Adv. Mater. Technol.* 2201557 (2023) doi:10.1002/ADMT.202201557.

47. Hoeng, F., Denneulin, A. & Bras, J. Use of nanocellulose in printed electronics: a review. *Nanoscale* **8**, 13131–13154 (2016).

48. Xu, Z., Wu, M., Gao, W. & Bai, H. A Transparent, Skin-Inspired Composite Film with Outstanding Tear Resistance Based on Flat Silk Cocoon. *Adv. Mater.* **32**, 2002695

(2020).

49. Sliz, R., Karzarjeddi, M., Liimatainen, H. & Fabritius, T. Nanocellulose as Sustainable Replacement for Plastic Substrates in Printed Electronics Applications. *Proc. 2021 IEEE 11th Int. Conf. "Nanomaterials Appl. Prop. N. 2021* (2021) doi:10.1109/NAP51885.2021.9568541.

50. Islam, A., Hansen, H. N., Tang, P. T. & Sun, J. Process chains for the manufacturing of molded interconnect devices. *Int. J. Adv. Manuf. Technol.* **42**, (2009).

51. Reis Carneiro, M., Majidi, C. & Tavakoli, M. Dielectric Elastomer Actuators with Biphasic Ag-EGaIn Electrodes. (2021) doi:10.1002/adem.202100953.

52. Wang, X., Liu, X. & Schubert, D. W. Highly Sensitive Ultrathin Flexible Thermoplastic Polyurethane/Carbon Black Fibrous Film Strain Sensor with Adjustable Scaffold Networks. *Nano-Micro Lett.* **13**, (2021).

53. Chen, S. Y., Zhuang, R. Q., Chuang, F. S. & Rwei, S. P. Synthetic scheme to increase the abrasion resistance of waterborne polyurethane–urea by controlling micro-phase separation. *J. Appl. Polym. Sci.* **138**, (2021).

54. Wang, Z., Peng, C., Yliniemi, K. & Lundström, M. Recovery of High-Purity Silver from Spent Silver Oxide Batteries by Sulfuric Acid Leaching and Electrowinning. *ACS Sustain. Chem. Eng.* **8**, (2020).

55. Tavakoli, M. *et al.* 3R Electronics: Scalable Fabrication of Resilient, Repairable, and Recyclable Soft-Matter Electronics. (2022) doi:10.1002/adma.202203266.

56. Deac, L. M. Foodborne Illness a Dynamic, Everywhere Possible Emergency Field Today. *J. Clin. Rev. Case Reports* **5**, (2020).

57. Tavakoli, M., Benussi, C. & Lourenco, J. L. Single channel surface EMG control of advanced prosthetic hands: A simple, low cost and efficient approach. *Expert Syst. Appl.* (2017) doi:10.1016/j.eswa.2017.03.012.

58. Hamedi, M., Salleh, S. H., Astaraki, M. & Noor, A. M. EMG-based facial gesture recognition through versatile elliptic basis function neural network. *Biomed. Eng. Online* **12**, (2013).

59. Moin, A. *et al.* An EMG Gesture Recognition System with Flexible High-Density Sensors and Brain-Inspired High-Dimensional Classifier. in *Proceedings - IEEE International Symposium on Circuits and Systems* vols 2018-May (2018).

60. Rissanen, S. M., Koivu, M., Hartikainen, P. & Pekkonen, E. Ambulatory surface electromyography with accelerometry for evaluating daily motor fluctuations in Parkinson's disease. *Clin. Neurophysiol.* **132**, (2021).

61. Pasmanasari, E. D. & Pawitan, J. A. The potential of electromyography signals as markers to detect and monitor Parkinson's disease. *Biomed. Pharmacol. J.* **14**, (2021).

62. Kotov-Smolenskiy, A. M., Khizhnikova, A. E., Klochkov, A. S., Suponeva, N. A. & Piradov, M. A. Surface EMG: Applicability in the Motion Analysis and Opportunities for Practical Rehabilitation. *Human Physiology* vol. 47 (2021).

63. Youn, B. Y. *et al.* Digital biomarkers for neuromuscular disorders: A systematic scoping review. *Diagnostics* **11**, (2021).

64. Yanowitz, F. G. *Introduction to ECG Interpretation*. (Intermountain Healthcare, 2018).

65. Reis Carneiro, M., Majidi, C. & Tavakoli, M. Multi-Electrode Printed Bioelectronic Patches for Long-Term Electrophysiological Monitoring. *Adv. Funct. Mater.* (2022) doi:10.1002/adfm.202205956.

Graphical abstract

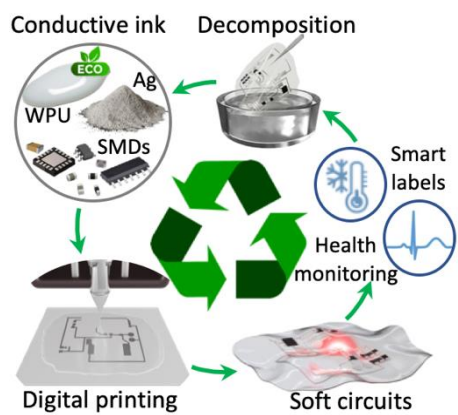